\documentclass[11pt,twoside]{article}

\usepackage{asp2006}
\usepackage{graphicx}
\usepackage{lscape}
\usepackage{multirow}
\usepackage[latin1]{inputenc}
\usepackage[OT1]{fontenc}

\newcommand{\AU}{\mbox{AU}}
\newcommand{\pc}{\mbox{pc}}
\newcommand{\mas}{\ensuremath{\mbox{mas}}}
\newcommand{\muas}{\ensuremath{\mu\mbox{as}}}
\newcommand{\Msun}{\ensuremath{\mbox{M}_{\rm Sun}}}
\newcommand{\Mearth}{\ensuremath{\mbox{M}_{\rm Earth}}}
\newcommand{\Mjupiter}{\ensuremath{\mbox{M}_{\rm Jupiter}}}

\bibpunct{(}{)}{;}{a}{}{,}    
\def\cite#1{\citealp{#1}}     

\makeatletter
\def\@jourvol{}
\def\cpr@year{2009}
\def\vol@title{Pathways Towards Habitable Planets}
\def\vol@author{V.\ Foresto, I.\ Ribas, D.\ Gelino (eds.)}
\makeatother

\begin{document}
\title{Detecting and characterizing extrasolar planetary systems with
  astrometry: review from the Blue Dots astrometry working
  group} 
\author{F.\ Malbet\altaffilmark{1}, A.\ Sozzetti\altaffilmark{2}, P.\ Lazorenko\altaffilmark{3}, R.\ Launhardt\altaffilmark{4}, D.\ Ségransan\altaffilmark{5}, F.\ Delplancke\altaffilmark{6}, N.\ Elias\altaffilmark{7}, M.\ Muterspaugh\altaffilmark{8}, A.\ Quirrenbach\altaffilmark{9}, S.\ Reffert\altaffilmark{9}, G.\ van Belle\altaffilmark{6}}
\altaffiltext{1}{Université de Grenoble J.~Fourier/CNRS, Laboratoire
  d'Astrophysique de Grenoble, BP 53, F-38041 Grenoble cedex 9,
  France. Email: \texttt{$<$Fabien.Malbet\@obs.ujf-grenoble.fr$>$}}    
\altaffiltext{2}{Osservatorio Astronomico di Torino INAF, Strada Osservatorio
  20, I-10025 Pino Torinese, Italy}    
\altaffiltext{3}{Main Astronomical Observatory, 27 Acad. Zabolotnoho St, UA Kiev 03680, Ukraine}    
\altaffiltext{4}{Max-Planck-Institut für Astronomie, Königstuhl 17, D-69117
  Heidelberg, Germany}    
\altaffiltext{5}{Observatoire astronomique de l'Université de Genève, Sauverny, CH-1290 Versoix, Switzerland}    
\altaffiltext{8}{Tennessee State University, 3500 John A Merritt Blvd,
  Nashville, TN 37209, U.S.A.}    
\altaffiltext{7}{National Radio Astronomy Observatory, 1003 Lopezville Road,
  Socorro NM 87801, U.S.A.}    
\altaffiltext{9}{Zentrum für Astronomie der Universität Heidelberg - Landessternwarte, Königstuhl 12, 69117 Heidelberg, Germany }    
\altaffiltext{6}{European Southern Observatory, Karl-Schwarzschild-Str. 2
D-85748 Garching bei München, Germany}    

\begin{abstract} 
  The astrometry technique is an important tool for detecting and
  characterizing exoplanets of different type. In this review, the
  different projects which are either operating, in construction or in
  discussion are presented and their performance discussed in the
  framework of the Blue Dots study. We investigate the
  sensitivity of astrometry to different sources of noise and we show
  that astrometry is a key technique in the path of discovering and
  characterizing new types of planets including the very challenging
  category of Earth-like planets orbiting the habitable zone of
  solar-type stars.
\end{abstract}



\section{Introduction}

In the Blue Dots study (see Foresto et al.\ in this volume) which aims
at preparing a road map towards the detection and characterization of
habitable exoplanets, precise stellar astrometry is one of the method
identified to detect exoplanets and characterize them. The goals of the
Astrometry working group within the Blue Dots initiative are to
identify the type of exoplanets that astrometry can detect and
characterize so that the Blue Dot table can be filled in with the
astrometry prospects, to investigate the limitations of astrometry and
finally emphasize the complementarity of astrometry with the other
techniques. The reader is invited to report to a deeper review on the
subject by \citet{2009arXiv0902.2063S} which is an essential resource for
this work.

\section{How astrometry can detect exoplanets?}




The motion of a star projected onto the plane of sky is a combination
of 3 types of apparent motion: parallax which is the apparent motion
due to the change of perspective of the observer (mainly the location
of the Earth during one year), the proper motion which is the motion
of the star+planets system in the galaxy, and the reflex motion due to
the presence of orbits. A precise orbit determination unravels the
presence of planets of different masses, only if one is able to
subtract the effect of parallax and proper motion.

It is generally assumed that a minimal signal-to-noise ratio of 5-6 on
the reflex motion of the star is required to detect a planet. In this
case, astrometric measurements of star motion yield the period $P$ and
the planetary mass $M_P$, but also the six parameters of the orbit: 
the semi-major axes $a_P$, the inclination of the orbit $i$, the
eccentricity of the orbit $e$, the longitude of the ascending node
$\Omega$, the argument of periapsis $\omega$ and the mean anomaly
$\nu$ at epoch $T_o$. 

\begin{table}[t]
  \centering
  \caption{Expected astrometric signal for different flavors of planetary systems.}
  \label{tab:astrometry_signal}
  \medskip
  \small
  \begin{tabular}{|p{3.1cm}|ccc|ccc|}
    \cline{2-7}
    \multicolumn{1}{c|}{}
    &\multicolumn{3}{|c}{\bf Giants planets}
    &\multicolumn{3}{|c|}{\bf Telluric planets} \\
    \hline
    \multirow{2}{*}{Type of planet} &Classical&Young   &Hot     &Hot        &Earth&Earth\\
                   &jupiter &jupiter&jupiter&super-Earth     &in HZ&in HZ \\
    Stellar spectral type   &G2 &G2 &G2 &M &G2 &M\\
    \hline
    $M_P$ (\Mearth) &300 &300 &300 &5 &1 &1 \\
    $M_P$ (\Mjupiter) &1 &1 &1 &0.02 &0.003 &0.003 \\
    $a_P$ (\AU) &5 &5 &0.1 &0.1 &1 &0.28 \\
    $P$ (yr) &11 &11 &0.03 &0.05 &1 &0.2 \\
    $P$ (d) &4084 &4084 &12 &17 &365 &82 \\
    $M_*$ (\Msun) &1 &1 &1 &0.45 &1 &0.45 \\
    $d$ (\pc) &10 &150 &10 &2.5 &10 &9 \\
    \hline
    \textbf{Astrometric}
    &\multirow{2}{*}{\textbf{495}} 
    &\multirow{2}{*}{\textbf{33}} 
    &\multirow{2}{*}{\textbf{10}} 
    &\multirow{2}{*}{\textbf{1}} 
    &\multirow{2}{*}{\textbf{0.3}}
    &\multirow{2}{*}{\textbf{0.2}} \\
    \textbf{signal} (in \muas) &&&&&& \\
    \hline
  \end{tabular}
\end{table}
The contribution of a planet to the reflex motion of its host star is given by the following formula: 
\begin{equation}
  \label{eq:1}
  \Delta \alpha = 0.33 \left(\frac{a_P}{1\,\AU}\right) 
  \left(\frac{M_P}{1\,\Mearth}\right)
  \left(\frac{M_*}{1\,\Msun}\right)^{-1} 
  \left(\frac{d}{10\,\pc}\right)^{-1} \muas
\end{equation}
Typical numbers corresponding to various flavors of planetary systems
are given in Table \ref{tab:astrometry_signal}. There are almost 4
orders of magnitude between a Jupiter in a Solar-like planetary system
located at $10\,\pc$ which gives a signal of the order of $500\,\muas$
and a Earth-like planets in its habitable zone which gives a signal of
$0.3\muas$. Astrometry, unlike some other methods, is best suited when
looking to nearby sources.

There are a variety of techniques to measure accurately the
astrometric motion of stars, i.e.\ to measure accurately the positions
of stars on the sky plane. These techniques can lead to wide-angle or
narrow-angle observations between stars, using relative or absolute
measurements, with local or global strategy. The atmosphere turbulence
is an important limitation to perform astrometry and therefore there
are both ground-based and space-born astrometric projects.

\section{Main astrometric projects}
\label{sec:main-astr-proj}

We have listed 5 main types of astrometric projects in the domain of
exoplanet detection and characterization. We have sorted them by
accuracy level and number of potential targets.

\subsection{Large ground-based telescopes}

For many years starting in the 1960's, a narrow-field astrometry was
frequently tried for finding exoplanets. These attempts resulted in
various discoveries, but none confirmed (Barnard's star, Lalande
21185). These failures caused by insufficient precision and systematic
errors of the photographic technique cast doubt on use of the
astrometry for exoplanet studies.  Besides, the narrow-field
astrometry was considered to be limited by a $\sim1\,\mas$ precision
due to the atmospheric image motion which came from the expression
$\varepsilon \sim \theta / D^{2/3}$ derived by
\citep{1980A&A....89...41L} for the r.m.s. of the image motion
$\varepsilon$ in the measurement of a distance $\theta$ between a pair
of stars on a telescope of diameter $D$.
Recently, \citet{2002A&A...382.1125L} has found that for the reference
field represented by a grid of stars, the power law is $\varepsilon
\sim \theta^{11/6} / D^{3/2}$ with an improvement with
$D$. Above expression is however asymptotic and refers to the very
narrow field mode of observations requesting use of very large
telescopes. In practice, this opens a way to a $<100\,\muas$
astrometry in a field of view  limited to $\theta \leq 1'$ with integration times of a
few minutes only, yet sufficient for the detection of massive
exoplanets around nearby stars. This challenging precision, of course,
assumes a proper handling of many other effects, in particular related
to a highly complicated shape of star profiles at the sub-pixel level
and to the unfixed, floating position of reference stars due to the
differential color refraction, proper motion, parallax, etc. Incorrect treatment of these
effects degrades precision to $\sim1\,\mas$. Currently, astrometric
search of exoplanets is renewed at three telescopes:
\begin{itemize}
\item The STEPS instrument installed on the 5-m Palomar telescope
  \citep{2004ApJ...617.1323P} with astrometric precision of $1\,\mas$
  per a single series of 20-30 exposures is used for exoplanet search
  around 30 late M stars since 1997. In 2009, this resulted in the
  first astrometric discovery of the giant exoplanet VB 10b around a
  main-sequence star \citep{2009ApJ...700..623P}, although it is still
  debated \citep{2009A&A...505L...5Z, 2009arXiv0912.0003B}.
\item The CAPSCam camera on the 2.5-m Las Campanas telescope has an
  estimated astrometric precision of $300\,\muas/\mbox{h}$. Search program
  includes about 100 late M, L, and T dwarfs to be observed for 10
  years \citep{2009PASP..121.1218B}.
\item The FORS2 camera installed on one of the 8-m VLT telescope was
  tested in a series of theoretical and observational studies which
  revealed its long-term astrometric precision of $50-100\,\muas$ at
  time scales from a few days to a few years
  \citep{2009A&A...505..903L}. Science observations started in 2009
  with the VB 10b as a program object and include 14 nearby L dwarfs.
\end{itemize}
Precision narrow-field astrometry is applicable to the targets of
12--17 mag at galactic latitudes up to $15-30^{\circ}$ allowing for a
sufficient number of reference stars. Astrometry enables the detection
of $\sim \Mjupiter$ planets at nearby low-mass red stars and brown dwarfs
with orbital period of $\geq 1$ year. We will take the range $0.3-1\,\mas$ as 
a conservative number for the accuracy of large telescope astrometry.

\subsection{Hubble Space Telescope  / Fine Guidance Sensors}

Relative, narrow-angle astrometry from space has been performed so far
with the \emph{Fine Guidance Sensors} aboard the \emph{Hubble Space
  Telescope} (HST/FGS). For HST/FGS astrometry with respect to a set
of reference objects near the target (within the $5\times5$\,arcsec
instantaneous field of view of FGS), $1-2\,\mas$ single measurement
precision down to $m_V\sim16$ has been demonstrated
\citep[see][]{1994PASP..106..327B, 1999AJ....118.1086B} using an ad
hoc calibration and data reduction procedures to remove a variety of
random and systematic error sources from the astrometric reference
frame (spacecraft jitter, temperature variations and
temperature-induced changes in the secondary mirror position, constant
and time-dependent optical field angle distortions, orbit
drifts, lateral color corrections). The limiting factor is the
spacecraft jitter. A single-measurement precision below $0.5-1\,\mas$
is out of reach for HST/FGS.  The first undisputed value of the actual
mass of a Doppler-detected planet was obtained by
\citet{2002ApJ...581L.115B} who derived the perturbation size,
inclination angle, and mass of the outer companion in the
multiple-planet system GJ\,876 from a combined fit to Hubble Space
Telescope/FGS astrometry and high-precision radial-velocities. Six
other objects are under observation.

\subsection{Ground-based dual star interferometry}

Long-baseline optical/infrared interferometry is an important
technique to obtain high-precision astrometry measurements
\citep{1992A&A...262..353S}. The idea is to operate an interferometer
equipped with a dual instrument observing two stars simultaneously so
that the optical delay between the fringes can be accurately measured.
The first observations have been achieved by the Palomar Testbed
Interferometer \citep[PTI, ][]{1999ApJ...510..505C} reaching $\sim100\,\muas$
short-term accuracy for $30''$ binaries \citep{2000Natur.407..485L} and
$20-50\,\muas$ for sub-arcsec binaries \citep{2004ApJ...601.1129L}
within the Palomar High-precision Astrometric Search for Exoplanet
Systems (PHASES) program.

Two projects are underway: the \emph{ASTrometric and phase-Referenced
  Astronomy}\citep[ASTRA:][]{2008arXiv0811.2264P} on the Keck
Interferometer and \emph{Phase-Referenced Imaging and Micro-arcsecond
  Astrometry} \citep[PRIMA:][]{2008NewAR..52..199D} at the VLT
interferometer.  These two projects are designed to perform
narrow-angle interferometric astrometry at an accuracy better than
$100\,\muas$ with two telescopes of a target and one reference star
separated by up to $1'$. This is however a challenging technique
requiring optical path difference errors less than $5\,\mbox{nm}$.

The Exoplanet Search with PRIMA (ESPRI) Consortium
\citep{2008SPIE.7013E..76L} foresees to carry out a three-fold observing
program focused on the astrometric characterization of known radial
velocity planets within $\leq 200\,\pc$ from the Sun, the astrometric
detection of low-mass planets around nearby stars of any spectral type
within $\approx15\,\pc$ from the Sun, and the search for massive planets
orbiting young stars with ages in the range $5-300\,\mbox{Myr}$ within
$\sim100\,\pc$ from the Sun. The target list includes $\sim100$
stars with good references. 

\subsection{Space-borne global astrometric Survey: Gaia}

In its all-sky survey, Gaia, due to launch in Spring 2012, will
monitor the astrometric positions of all point sources in
the magnitude range $6-20\,\mbox{mag}$, a database encompassing
$\sim10^9$ objects.  Using the continuous scanning principle first
adopted for Hipparcos, Gaia will determine positions, proper motions,
and parallaxes for all objects, with end-of-mission precision between
$6\,\muas$ at $V=6\,\mbox{mag}$ and $200\,\muas$ at $V=20\,\mbox{mag}$
and an averaged 80 transits per object in 5 years.

Gaia astrometry, complemented by on-board spectrophotometry and
radial velocity information, will have the precision
necessary to quantify the early formation, and subsequent dynamical,
chemical and star formation evolution of the Milky Way Galaxy. 
One of the relevant areas in which the Gaia observations will have
great impact is the astrophysics of planetary systems
\citep[e.g.][]{2008A&A...482..699C}, in particular when seen as a
complement to other techniques for planet detection and
characterization \citep[e.g.][]{2009arXiv0902.2063S}.

Using Galaxy models, our current knowledge of exoplanet frequencies,
and Gaia estimated precision $\sim 10\,\muas$ on bright targets ($V
< 13$), \citet{2008A&A...482..699C} have shown that Gaia will measure
actual masses and orbital parameters for possibly thousands of giant
planets and determine the degree of coplanarity of possibly hundreds
of multiple-planet systems. Gaia will be sensitive as far as the
nearest star-forming regions for systems with massive giant planets on
$1\leq a \leq4\,\AU$ orbits around solar-type hosts and out to
$30\,\pc$ for Saturn-mass planets with similar orbital semi-major axes
around late-type stars.

Gaia holds promise for crucial contributions to many aspects of
planetary systems astrophysics, in combination with present-day and
future extrasolar planet search programs. Gaia data over the next
decade will allow us to (a) significantly refine of our understanding
of the statistical properties of extrasolar planets, (b) carry out
crucial tests of theoretical models of gas giant planet formation and
migration, (c) achieve key improvements in our comprehension of
important aspects of the formation and dynamical evolution of
multiple-planet systems, (d) provide important contributions to the
understanding of direct detections of giant extrasolar planets, and
(e) collect essential supplementary information for the optimization
of the target lists of future observatories aiming at the direct
detection and spectroscopic characterization of terrestrial, habitable
planets in the vicinity of the Sun.

\subsection{Space-based astrometric observatory: SIM-Lite}

The \emph{Space Interferometry Mission} is a space borne instrument
\citep{2008SPIE.7013E..79M} which would carry out astrometry to
micro-arc second precision on the visible light from a large sample of
stars in our galaxy and search for earth-like planets around nearby
stars \citep{2008PASP..120...38U}. SIM-Lite is an alternative concept
for SIM and it is the current proposed implementation for SIM for
NASA program on Search for Earth-like planets and life. The SIM-Lite
instrument is an optical interferometer with a baseline of 6 meters
which include a guiding interferometer and a guiding siderostat for
spacecraft pointing and a Science interferometer to perform high
accuracy astrometric measurements on target stars.  The primary
objective for SIM-Lite is to search 65 nearby stars for exoplanets of
masses down to one Earth mass, in the Habitable Zone.  SIM-Lite is
designed to deliver better than $1\,\muas$ narrow-angle astrometry in
$1.5\,\mbox{hr}$ integration time on bright targets ($m_V \leq 7$) and
moderately fainter references ($m_v\simeq 9-10$)
\citep{2008SPIE.7013E.151G}.  An accuracy on the position of the delay
lines of a few tens of picometers with a 6-m baseline must be achieved
\citep{2008SPIE.7013E.157Z}. Furthermore, a positional stability of
internal optical path lengths of $\sim 10\,\mbox{nm}$ is required, in order to
ensure maintenance of the fringe visibility.

SIM-Lite has planned for 3 planetary system survey.  (1) The
\emph{deep planetary survey} will focus on less than one hundred
nearby stars of the main sequence within 10 parsecs from the Sun. The
main objective is to identify planetary system with Earth-like planets
in the habitable zone around these Sun-like stars. This deep survey
requires the highest possible astrometric accuracy, below the single
micro-arc-second. This accuracy is achieved by multiple visits to the
target stars with between 12 to 60 chops per visits in order to lower
the instrument accuracy down to the required level for detection of
one Earth mass planet. A 60 chop visit to a magnitude 6 target will
require less than two hours of observation time. The target would then
be observed several hundred times during the lifetime of the
mission. (2) The \emph{broad planetary survey} will study more than
one thousand set of stars of many types including O, B, A, F, binary
with a reduced astrometric accuracy of $5\,\muas$ in order to cover
the diversity of planetary systems and to increase our knowledge on
the nature and the evolution of planetary systems in their full
variety.  This accuracy can be achieved by short visits to the target
stars. To determine accurately the orbit parameters of the planets,
100 visits per target star will be scheduled over the 5 year
mission. (3) The \emph{ young planetary system survey} will observe
less than one hundred nearby solar type stars with ages below 100
millions years within 100 parsecs from the Sun with the aim to
understand the frequency of Jupiter-mass planets and the early
dynamical evolution of planetary systems.  This survey can be conducted
with a reduced astrometric accuracy of $4\,\muas$ achieved with
4-chops visits requiring about eight minutes of observation time for a
$V\sim11$ young star.

\section{Limitations and performance}

Using astrometry to detect and characterize extra-solar planetary
systems is challenging in many aspects. In this part we focus on the
main issues that astrometry will have to face especially in the field
of Earth-mass planets.

\subsection{Influence of giant planets on Earth-like planets detection}

Correct determination of the astrometric orbits of planetary systems
involves highly non-linear orbital fitting procedures, with a large
number of model parameters. Particular attention will have to be
devoted, for example, to the assessment of the relative robustness and
reliability of different procedures for orbital fits (and associated
uncertainties).  The quality of the coplanarity analysis for multiple
systems has to be measured against the achieved
single-measurement precision and available redundancy in the
observations. Correctly identifying
signals with amplitude close to the measurement uncertainties is a challenge
particularly in the presence of larger signals induced by other
companions and/or sources of astrophysical noise of comparable
magnitude, like Jupiter-like planets  for telluric ones ($500\,\muas$
signal compared to $0.3\,\muas$ signal in the case of a Solar System
located at $10\,\pc$.

All these issues could have a significant impact on the capability of
Gaia and SIM-Lite to detect and characterize planetary systems.
Double-blind test campaigns have been carried out to estimate the
potential of both Gaia and SIM-Lite for detecting and measuring
planetary systems \citep{2008A&A...482..699C, 2009arXiv0904.0822T}.
The Gaia \emph{Data Processing and Analysis Consortium} (DPAC) is
developing the modeling of the astrometric signals produced by
planetary systems implementing multiple robust procedures for
astrometric orbit fitting (such as Markov Chain Monte Carlo and
genetic algorithms) and in order to determine the degree of dynamical
stability of multiple-component systems. A similar work is also
underway for SIM-Lite (see Traub et al.\ in this volume) which shows
that detection completeness is 90\% for all planets reaching 95\% for
terrestrial planets in the habitable zone and a reliability of
respectively 97\% and 100\%.

\subsection{Impact of stellar activity}

In the domain of terrestrial planets, the expected signal is so small
($0.3\muas$ for an Earth at $10\,pc$) that one must take into account
the effect of the presence of spots on the surface of the target
stars. These spots will introduce a spurious signal the level of which
should be evaluated all the more beacuse lifetime of these spots can last
several weeks. This phenomenon also impacts the other detection
techniques like transits and radial velocity measurements. Several
works have addressed this issue in the case of astrometry at least for
the most sensitive instruments, Gaia and SIM-Lite
\citep{2007A&A...476.1389E, 2009A&A...506.1469D, 2009arXiv0911.2008M}.

The general conclusion is that, unlike the radial-velocity case,
\muas-level astrometry is significantly less affected by the above
astrophysical noise sources. For example, a Sun-like star inclined at $i=90^\circ$ at 10\,pc is predicted to have a jitter of $0.087\,\muas$ in its astrometric
position along the equator below the expected signal of an Earth at a
level of $0.3\,\muas$, and $0.38\,\mbox{m/s}$ in radial velocities
  above the expected signal from a Earth at $0.09\,\mbox{m/s}$. If
the presence of spots due to stellar activity is the ultimate limiting
factor for planet detection, then the sensitivity of SIM Lite to Earth-like
planets in habitable zones is about an order of magnitude higher that
the sensitivity of prospective ultra-precise radial velocity
observations of nearby stars.

\subsection{Performance and mission scales}

\begin{table}[t]
  \centering
  \caption{Summary of the astrometry mission scopes and their scales.}
  \label{tab:scales}
  \smallskip
  \includegraphics[width=\hsize]{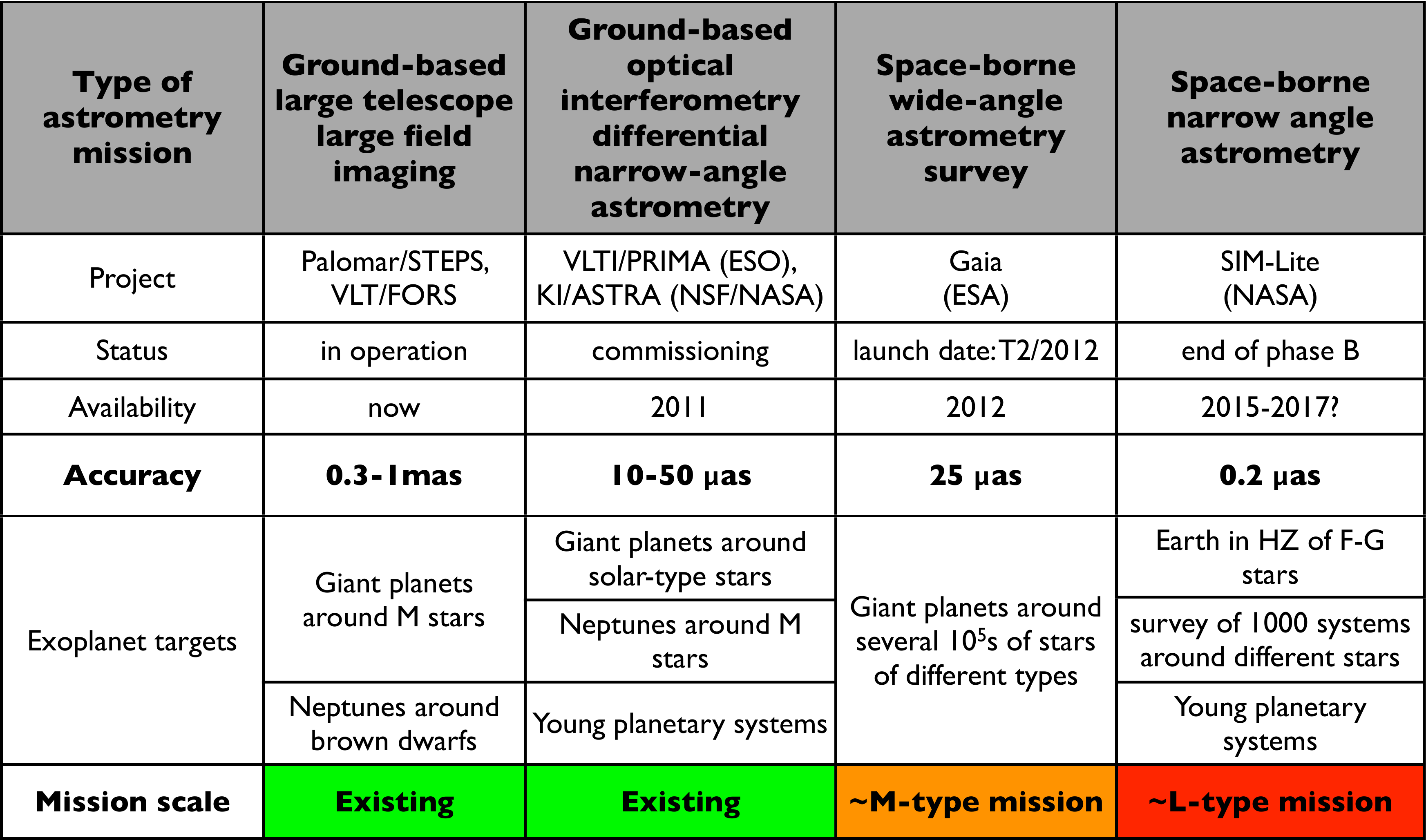}
\end{table}

Table \ref{tab:scales} summarizes the various projects presented in
Sect.\ \ref{sec:main-astr-proj} with their status, since when they are
operating or when they will happen, the expected accuracy and their main
targets. We have also indicated their scale in the \emph{Blue Dots}
terminology (see Foresto et al.\ in this volume). Ground-based
facilities are categorized in existing low cost projects whereas Gaia
is in the M-type mission class and SIM-Lite in the L-type mission even
though if approved it will fly and operate earlier than some
ground-based instruments (like spectrographs on extremely large
telescopes).

\section{Conclusion: specificity of astrometry in the exoplanet field}

This review has allowed us to show that astrometry can detect and
characterize new categories of exoplanets especially in the Solar
neighborhood. It encompasses techniques ranging from ground-based
instruments limited to the largest planets to space borne missions
allowing Earth-like planets to be detected in the habitable zone
around solar-type stars. Astrometry will obtain full
characterization of orbits which gives us (a) masses and tells us (b)
where and when to look with spectroscopic characterization missions.

Astrometry is mostly immune to stellar activity even at the signal
level due to Earth-like planets and therefore is well positioned for
identifying and characterizing planets for future direct detection and
spectroscopic follow-up projects like DARWIN/TPFI or TPF-c.

Astrometry is a difficult and challenging technique, yet it may be
the only way to detect and characterize Earth-like planets in the
habitable zone of solar-type stars. Last but not the least, astrometry
projects are underway or technically ready and should soon contribute
significantly to the exoplanet field.

\acknowledgements 
We would like to thank S.\ Shaklan, S.\ Pravdo, F.\ Mignard, M.\ Shao,
J.\ Catanzarite, V.\ Makarov for providing material for the review
presentation in Barcelona meeting.


\end{document}